\documentclass[conference]{IEEEtran}
\IEEEoverridecommandlockouts
\usepackage{cite}
\usepackage{amsmath,amssymb,amsfonts}
\usepackage{algorithmic}
\usepackage{graphicx}
\usepackage{textcomp}
\usepackage{xcolor}
\usepackage{balance}
\usepackage{cite}
\usepackage{hyperref}
\usepackage{amsmath,amssymb,amsfonts}
\usepackage{amsmath}
\usepackage{amsthm}

\usepackage{graphicx}
\usepackage{textcomp}
\usepackage{xcolor}
\usepackage{graphicx}
\usepackage{float}
\usepackage{subfigure}
\usepackage{amsmath}
\usepackage{amsfonts,amssymb}
\usepackage{mathrsfs}
\usepackage{mathtools}
\usepackage{bm}
\usepackage{multirow}
\usepackage{array}
\usepackage{amssymb}
\usepackage{amsmath}
\usepackage{cite}
\usepackage{url}
\usepackage{xcolor}
\usepackage{cite,graphicx,amsmath,amssymb}
\usepackage{subfigure}
\usepackage{fancyhdr}
\usepackage{mdwmath}
\usepackage{mdwtab}
\usepackage{caption}
\usepackage{amsthm}
\usepackage{setspace}
\usepackage{bm}
\usepackage{mathtools}
\usepackage{dsfont}
\usepackage{bbm}
\newtheorem{remark}{Remark}
\newtheorem{theorem}{Theorem}

\newtheorem{lemma}{Lemma}

\newtheorem{corollary}{Corollary}

\makeatletter
\newcommand{\biggg}{\bBigg@{3}}
\newcommand{\Biggg}{\bBigg@{3.5}}
\makeatother
\def\BibTeX{{\rm B\kern-.05em{\sc i\kern-.025em b}\kern-.08em
    T\kern-.1667em\lower.7ex\hbox{E}\kern-.125emX}}
\begin{document}

\title{Some Discussions on PHY Security in DF Relay}

\author{\IEEEauthorblockN{Chongjun Ouyang, Hao Xu, Xujie Zang, and Hongwen Yang}
School of Information and Communication Engineering\\
Beijing University of Posts and Telecommunications, Beijing, 100876, China\\
\{DragonAim, Xu\_Hao, zangxj, yanghong\}@bupt.edu.cn}

\maketitle

\begin{abstract}
Physical layer (PHY) security in decode-and-forward (DF) relay systems is discussed. Based on the types of wiretap links, the secrecy performance of three typical secure DF relay models is analyzed. Different from conventional works in this field, rigorous derivations of the secrecy channel capacity are provided from an information-theoretic perspective. Meanwhile, closed-form expressions are derived to characterize the secrecy outage probability (SOP). For the sake of unveiling more system insights, asymptotic analyses are performed on the SOP for a sufficiently large signal-to-noise ratio (SNR). The analytical results are validated by computer simulations and are in excellent agreement.
\end{abstract}


\section{Introduction}\label{sec1}
Current wireless communication systems are generally composed of multi-terminal networks, and a fundamental building block of this topology is the source-relay-destination modality \cite{Nosratinia2004,Hu2018}. This modality is especially welcome when the destination is far away from the source and the direct link therein is blocked by deep fading. Typical relaying protocols include amplify-and-forward (AF) and decode-and-forward (DF). Amplify-and-forward, as its name suggests, means that the relay node amplifies the received signals sent by the source and then forwards them to the destination. As for the decode-and-forward, the relay node will decode the signals from the source before forwarding them to the destination, hence yielding a better performance than the AF relay \cite{Nosratinia2004}. By now, these two protocols have been regarded as two effective strategies to resist the deep fading involved in long-haul wireless transmission \cite{Nosratinia2004,Hu2018}.

In addition to the severe fading, wireless networks are also fragile to malicious eavesdropping and the accompanying security issues. More specifically, because of the open access property of wireless medium, malicious eavesdroppers can intercept the signal transmission from source to its intended destination. This spawned the development and application of physical layer (PHY) security where wiretap or secrecy coding is exploited to ensure perfect security, i.e., the eavesdroppers cannot decipher or decode any confidential information from the wiretapped messages \cite{Wyner1975,Cheong1978,Wu2018}.

Given this backdrop, the secrecy performance of relay systems has been discussed for more than twenty years \cite{Oohama2001,Zou2016,Zou2013,Kundu2015,Yang2017,Zhao2017,Guo2017,Lei2017,Lei2019,Zhao2019,Li2021}. Among the numerous research efforts on this topic, the wiretap models of AF and DF relay systems and the relative analyses of secrecy channel capacity, established by Zou \emph{et al.}, are most widely accepted \cite{Zou2013}. However, although this treatise has led to a proliferation of studies in this field \cite{Kundu2015,Yang2017,Zhao2017,Guo2017,Lei2017,Lei2019,Zhao2019,Li2021}, the discussion on the secrecy channel capacity of DF systems in \cite{Zou2013} lacks rigorousness. As a result, the derived expression therein cannot describe the exact secrecy capacity of the secure DF system. As known to all, the secrecy capacity of wiretap channels is the cornerstone of PHY security, which defines the upper limit of the error-free as well as perfectly secure information transmission rate \cite{Cheong1978}. Thus, using this inexact expression to analyze the PHY security of DF systems will undoubtedly weaken the practical significance of the developed analytical results. Nevertheless, it is worth noting that many researchers based their discussions precisely on this inexact expression \cite{Kundu2015,Yang2017,Zhao2017,Guo2017,Lei2017,Lei2019,Zhao2019,Li2021}.

To remedy this theoretical flaw, this paper investigates the PHY security in DF relay systems. Specifically, we consider three typical models of secure DF systems and provide rigorous discussions on the secrecy capacity therein. These three models are categorized by the types of wiretap links, which also include the widely discussed model in conventional works \cite{Zou2013,Kundu2015,Yang2017,Zhao2017,Guo2017,Lei2017,Lei2019,Zhao2019,Li2021}. Based on the newly formulated expressions of the secrecy channel capacity, closed-form analytical results are developed to characterize an important secrecy performance evaluation metric, i.e., the secrecy outage probability (SOP). To obtain an insightful understanding of the secure DF relay system, we also perform asymptotic analyse on the SOP in the high signal-to-noise ratio (SNR) region and specify the secrecy diversity order (SDO).

The remaining parts of this paper is structured as follows: Section \ref{sec2} describes the system model. In Section \ref{sec4}, the SOPs of all the considered secure DF relay cases are investigated. The numerical results and corresponding analysis are shown in Section \ref{sec5}. Finally, Section \ref{sec5} concludes the paper.

\begin{figure*}[!t]
    \centering
    \subfigbottomskip=0pt
	\subfigcapskip=0pt
\setlength{\abovecaptionskip}{0pt}
   \subfigure[Case \uppercase\expandafter{\romannumeral1}.]
    {
        \includegraphics[height=0.11\textwidth]{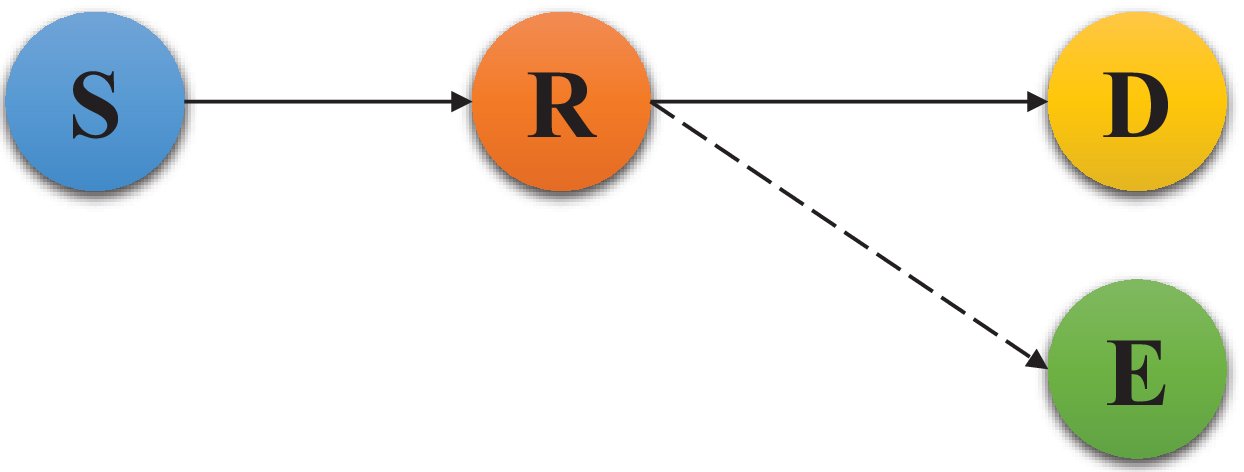}
	   \label{System_Model1}	
    }
    \subfigure[Case \uppercase\expandafter{\romannumeral2}.]
    {
        \includegraphics[height=0.11\textwidth]{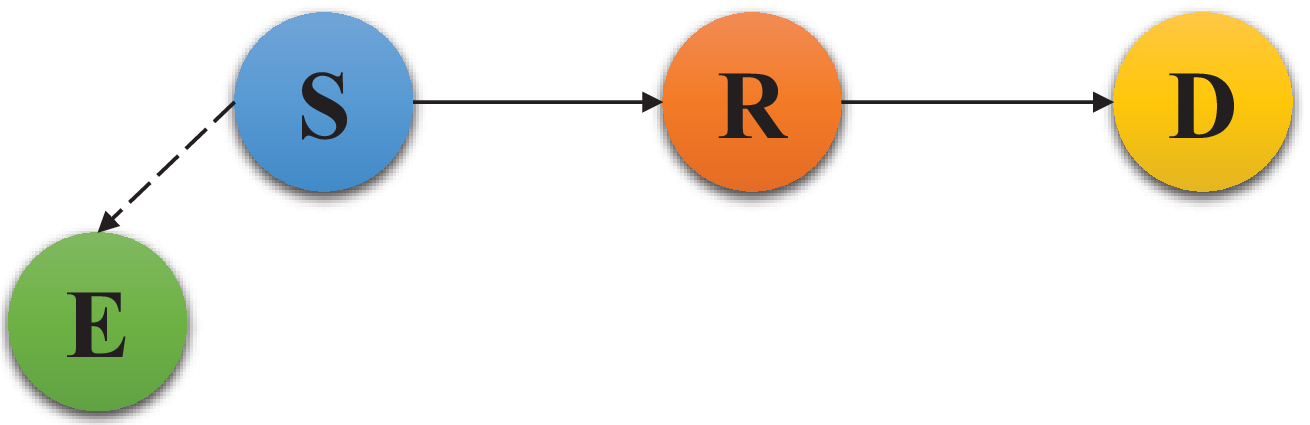}
	   \label{System_Model2}	
    }
    \subfigure[Case \uppercase\expandafter{\romannumeral3}.]
    {
        \includegraphics[height=0.11\textwidth]{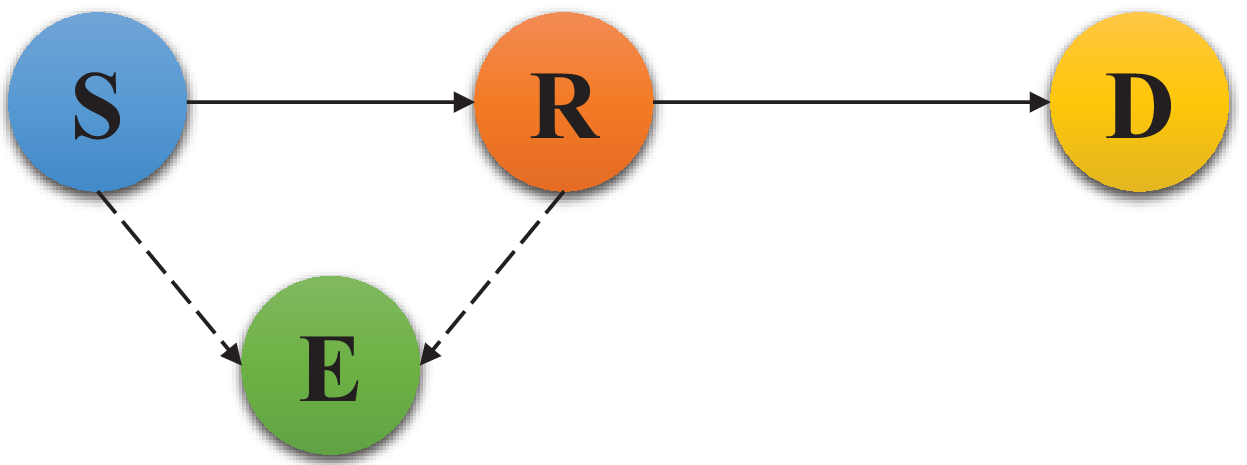}
	   \label{System_Model3}	
    }
\caption{Illustration of secure DF relay systems.}
    \label{figure1}
\end{figure*}

\section{System Model}\label{sec2}
There is a system consisting of one source (S), one destination (D), and one decode-and-forward (DF) relay (R) in the presence of an eavesdropper (E) as shown in {\figurename} {\ref{figure1}}, where all nodes are single-antenna devices and the solid and dash lines represent the main and wiretap links, respectively. The channel gain of the $i$-$j$ ($i,j\in\left\{{\text{S}},{\text{R}},{\text{D}},{\text{E}}\right\}$) link, denoted by $h_{i,j}$, is modeled as $h_{i,j}\sim\mathcal{CN}\left(0,1\right)$ (namely Rayleigh fading channel) and the thermal noise received at node $j$ ($j\in\left\{{\text{D}},{\text{R}},{\text{E}}\right\}$) is modeled as $\mathcal{CN}\left(0,\sigma_j^2\right)$ with $\sigma_j^2$ being the noise power. Throughout this work, we consider that the direct link between S and D is blocked by deep fading and the relay works in a DF mode. On this condition, S firstly sends the confidential information to R and then R decodes this message and forwards it to D. Furthermore, for the sake of fairness, we consider that the total transmit power, $P$, is allocated equally to S and R, respectively.

Before further discussions, we first review the secrecy channel capacity in the DF relay channel. By definition, the secrecy capacity is the upper limit of the secrecy coding rate, below which error-free as well as perfectly secure information transmission is possible. In DF relay channel, the secrecy coding can be exploited at both the source node and the relay node. Since the same confidential information is transmitted over the S-R and R-D links, the secrecy capacity of the entire DF relay channel can be written as
\begin{align}\label{Expression_Secrecy_Capacity_DF_Relay}
\mathcal{R}_{\rm{s}}=\max\{\min\left\{\mathcal{R}_{{\text{R}}\rightarrow{\text{D}}},\mathcal{R}_{{\text{S}}\rightarrow{\text{R}}}\right\},0\},
\end{align}
where $\mathcal{R}_{{\text{S}}\rightarrow{\text{R}}}$ and $\mathcal{R}_{{\text{R}}\rightarrow{\text{D}}}$ represent the secrecy capacities of the confidential information transmitting from S to R and from R to D, respectively. Based on this, we now intend to investigate three typical secure DF relay cases depending on the types of wiretap links.

\subsection{Case \uppercase\expandafter{\romannumeral1}}\label{sec3a}
For the first case shown in {\figurename} {\ref{System_Model1}}, the eavesdropper can only overhear the confidential information dedicated to D from R. We comment that this case has been widely discussed on current literature; see \cite{Zou2013,Kundu2015,Yang2017,Zhao2017,Guo2017,Lei2017,Lei2019,Zhao2019,Li2021} and the references therein. In this case, S does not need to utilize secrecy coding, and conventional coding methods are enough for reliable as well as secure communications. Yet, since E can receive the wireless signals from R, it is necessary for R to exploit secrecy coding in order to prevent E from knowing any useful information from its received signal. The secrecy capacity in the R-D link is given by \cite{Cheong1978}
\begin{align}
\mathcal{R}_{{\text{R}}\rightarrow{\text{D}}}^{1}=\max\left\{\log_2\left({\frac{1+{\bar\gamma}_{\text{D}}\left|h_{{\text{R}},{\text{D}}}\right|^2}
{1+{\bar\gamma}_{\text{E}}\left|h_{{\text{R}},{\text{E}}}\right|^2}}\right),0\right\},
\end{align}
where ${\bar\gamma}_{\text{D}}=\frac{P}{2\sigma_{\text{D}}^2}$ and ${\bar\gamma}_{\text{E}}=\frac{P}{2\sigma_{\text{E}}^2}$. It is worth mentioning that since the eavesdropper cannot get access to the wireless signals sent by S, the secrecy capacity in the S-R link equals the channel capacity of this link. Therefore, the secrecy capacity or the channel capacity in the S-R link is given by $\mathcal{R}_{{\text{S}}\rightarrow{\text{R}}}^{1}=\log_2\left(1+{\bar\gamma}_{\text{R}}\left|h_{{\text{S}},{\text{R}}}\right|^2\right)$, where ${\bar\gamma}_{\text{R}}=\frac{P}{2\sigma_{\text{R}}^2}$. Taken together, the secrecy capacity from S to D is given by
\begin{align}
\mathcal{R}_1=\max\left\{\min\left\{\mathcal{R}_{{\text{R}}\rightarrow{\text{D}}}^{1},\mathcal{R}_{{\text{S}}\rightarrow{\text{R}}}^{1}\right\},0\right\}.
\end{align}
Recall that the commonly used expression for the secrecy capacity of Case \uppercase\expandafter{\romannumeral1} is given as \cite{Zou2013,Kundu2015,Yang2017,Zhao2017,Guo2017,Lei2017,Lei2019,Zhao2019,Li2021}
\begin{align}\label{Expression_Old_Secrecy_Capacity}
\hat{\mathcal{R}}_1=\max\left\{\log_2\left({\frac{1+\min\left\{{\bar\gamma}_{\text{D}}\left|h_{{\text{R}},{\text{D}}}\right|^2,{\bar\gamma}_{\text{R}}\left|h_{{\text{S}},{\text{R}}}\right|^2
\right\}}{1+{\bar\gamma}_{\text{E}}\left|h_{{\text{R}},{\text{E}}}\right|^2}}\right),0\right\}.
\end{align}
We note that \eqref{Expression_Old_Secrecy_Capacity} can be rewritten as
\begin{align}\label{Expression_Old_Secrecy_Capacity_Type2}
\hat{\mathcal{R}}_1=\max\left\{\min\left\{\mathcal{R}_{{\text{R}}\rightarrow{\text{D}}}^{1},
\log_2\left({\frac{1+{\bar\gamma}_{\text{R}}\left|h_{{\text{S}},{\text{R}}}\right|^2}
{1+{\bar\gamma}_{\text{E}}\left|h_{{\text{R}},{\text{E}}}\right|^2}}\right)\right\},0\right\}.
\end{align}

It is clear that $\mathcal{R}_1\neq \hat{\mathcal{R}}_1$. Based on \eqref{Expression_Secrecy_Capacity_DF_Relay}, we note that the expression in \eqref{Expression_Old_Secrecy_Capacity_Type2} (or \eqref{Expression_Old_Secrecy_Capacity}) is obtained by treating $\log_2\left({\frac{1+{\bar\gamma}_{\text{R}}\left|h_{{\text{S}},{\text{R}}}\right|^2}
{1+{\bar\gamma}_{\text{E}}\left|h_{{\text{R}},{\text{E}}}\right|^2}}\right)$ as the secrecy capacity of the S-R link, which looks weird and makes no sense. On the one hand, the S-R link suffers no eavesdropping and the secrecy channel capacity of this link degenerates into its channel capacity, namely $\mathcal{R}_{{\text{S}}\rightarrow{\text{R}}}^{1}=\log_2\left(1+{\bar\gamma}_{\text{R}}\left|h_{{\text{S}},{\text{R}}}\right|^2\right)$ rather than $\mathcal{R}_{{\text{S}}\rightarrow{\text{R}}}^{1}=\log_2\left({\frac{1+{\bar\gamma}_{\text{R}}\left|h_{{\text{S}},{\text{R}}}\right|^2}
{1+{\bar\gamma}_{\text{E}}\left|h_{{\text{R}},{\text{E}}}\right|^2}}\right)$. On the other hand, the term $\log_2\left({\frac{1+{\bar\gamma}_{\text{R}}\left|h_{{\text{S}},{\text{R}}}\right|^2}
{1+{\bar\gamma}_{\text{E}}\left|h_{{\text{R}},{\text{E}}}\right|^2}}\right)$ represents the difference of channel capacities of the S-R and R-E links, which is of no physical meaning to the secure DF relay system. Therefore, the term $\log_2\left({\frac{1+{\bar\gamma}_{\text{R}}\left|h_{{\text{S}},{\text{R}}}\right|^2}
{1+{\bar\gamma}_{\text{E}}\left|h_{{\text{R}},{\text{E}}}\right|^2}}\right)$ should not appear in the expression of the secrecy capacity of Case \uppercase\expandafter{\romannumeral1}. Taken together, one can see that the expression in \eqref{Expression_Old_Secrecy_Capacity} (or \eqref{Expression_Old_Secrecy_Capacity_Type2}) makes less practical sense from a rigorous view of information theoretic secrecy \cite{Cheong1978}.

\subsection{Case \uppercase\expandafter{\romannumeral2}}\label{sec3b}
We then consider another case shown in {\figurename} {\ref{System_Model2}}, where the eavesdropper can only overhear the confidential information from S. In this case, it is necessary for S to exploit secrecy coding in the transmission over the S-R link, whereas it is enough for R to exploit conventional coding methods. Under this circumstance, the secrecy capacity in the S-R link is given as $\mathcal{R}_{{\text{S}}\rightarrow{\text{R}}}^{2}=\log_2\left({\frac{1+{\bar\gamma}_{\text{R}}\left|h_{{\text{S}},{\text{R}}}\right|^2}
{1+{\bar\gamma}_{\text{E}}\left|h_{{\text{S}},{\text{E}}}\right|^2}}\right)$, whereas the secrecy capacity or the channel capacity of the R-D link is given as $\mathcal{R}_{{\text{R}}\rightarrow{\text{D}}}^{2}=\log_2\left({1+{\bar\gamma}_{\text{D}}\left|h_{{\text{R}},{\text{D}}}\right|^2}\right)$. Hence, the secrecy capacity of the whole system is given as
\begin{align}
\mathcal{R}_2=\max\left\{\min\left\{\mathcal{R}_{{\text{S}}\rightarrow{\text{R}}}^{2},\mathcal{R}_{{\text{R}}\rightarrow{\text{D}}}^{2}\right\},0\right\}.
\end{align}

\subsection{Case \uppercase\expandafter{\romannumeral3}}\label{sec3c}
Finally, let us move to the case shown in {\figurename} {\ref{System_Model3}}, where E can overhear the confidential information from both S and R. Under this circumstance, it is necessary for both S and R to exploit secrecy coding to convey the information. Under this circumstance, the secrecy capacity in the S-R link is given as $\mathcal{R}_{{\text{S}}\rightarrow{\text{R}}}^{3}=\log_2\left({\frac{1+{\bar\gamma}_{\text{R}}\left|h_{{\text{S}},{\text{R}}}\right|^2}
{1+{\bar\gamma}_{\text{E}}\left|h_{{\text{S}},{\text{E}}}\right|^2}}\right)$, whereas the secrecy capacity of the R-D link is given as $\mathcal{R}_{{\text{R}}\rightarrow{\text{D}}}^{3}=\log_2\left({\frac{1+{\bar\gamma}_{\text{D}}\left|h_{{\text{R}},{\text{D}}}\right|^2}
{1+{\bar\gamma}_{\text{E}}\left|h_{{\text{R}},{\text{E}}}\right|^2}}\right)$. Moreover, we assume that different ensembles of secrecy codes are adopted by S and R. Hence, the secrecy capacity of the whole system is given as
\begin{align}
\mathcal{R}_3=\max\left\{\min\left\{\mathcal{R}_{{\text{S}}\rightarrow{\text{R}}}^{3},\mathcal{R}_{{\text{R}}\rightarrow{\text{D}}}^{3}\right\},0\right\}.
\end{align}

It is worth noting that the secrecy performance in Case \uppercase\expandafter{\romannumeral1} has been widely investigated (though with limited rigorousness and room for improvement) \cite{Zou2013,Kundu2015,Yang2017,Zhao2017,Guo2017,Lei2017,Lei2019,Zhao2019,Li2021}, whereas the secrecy performance in Case \uppercase\expandafter{\romannumeral2} and Case \uppercase\expandafter{\romannumeral3} has received less research attention. Due to this fact, the objective of this work is to provide a thorough as well as rigorous discussion on the secrecy performance achieved in all the aforementioned cases.

\section{Secrecy Performance Analyses}\label{sec4}
Having introduced the three cases and given the corresponding expressions of the secrecy capacity, we now move on to evaluating the secrecy performance of these three cases by analyzing the secrecy outage probability. The SOP is defined as the probability that the secrecy capacity is smaller than a preset value $\mathcal{R}>0$, i.e.,
\begin{align}
\mathcal{P}_i=\Pr\left(\mathcal{R}_i<\mathcal{R}\right),i=1,2,3.
\end{align}
\subsection{Case \uppercase\expandafter{\romannumeral1}}\label{sec4a}
Particularly, the following theorem gives the exact expression of the SOP.
\vspace{-5pt}
\begin{theorem}\label{Theorem_Case1_1}
The SOP under Case \uppercase\expandafter{\romannumeral1} is given as
\begin{align}\label{Theorem_Case1_1_Exp}
\mathcal{P}_1=1-{\rm{e}}^{-\frac{2^{\mathcal{R}}-1}{\bar\gamma_{\text{R}}}-\frac{2^{\mathcal{R}}-1}{\bar\gamma_{\text{D}}}}
\frac{1}{1+\frac{2^{\mathcal{R}}}{\bar\gamma_{\text{D}}}{\bar\gamma_{\text{E}}}}.
\end{align}
\end{theorem}
\vspace{-5pt}
\begin{IEEEproof}
Using the statistical independency of $\mathcal{R}_{{\text{R}}\rightarrow{\text{D}}}$ and $\mathcal{R}_{{\text{S}}\rightarrow{\text{R}}}$, we can calculate the SOP of Case \uppercase\expandafter{\romannumeral1} as
\begin{align}
\mathcal{P}_1=1-\Pr{\left({\mathcal{R}_{{\text{R}}\rightarrow{\text{D}}}^{1}}>{\mathcal{R}}\right)}
\Pr{\left({\mathcal{R}_{{\text{S}}\rightarrow{\text{R}}}^{1}}>{\mathcal{R}}\right)}.
\end{align}
Note that $\left|h_{{\text{R}},{\text{D}}}\right|^2$, $\left|h_{{\text{R}},{\text{E}}}\right|^2$, and $\left|h_{{\text{S}},{\text{R}}}\right|^2$ are all exponential distributed. Leveraging this fact, we can obtain the final results in \eqref{Theorem_Case1_1_Exp} after some basic mathematical manipulations.
\end{IEEEproof}
To gain more insights, we perform asymptotic analyses to the SOP for a sufficiently large signal-to-noise ratio. More specifically, we study two typical scenarios: 1) the SNR of the S-R link tends to infinity, whereas the SNR of the wiretap link is fixed; 2) the SNRs of both the S-R and the wiretap links tend to infinity. Particularly, the follow theorem describes the asymptotic SOP in the first scenario.
\vspace{-5pt}
\begin{theorem}\label{Theorem_Case1_2}
For fixed $\bar\gamma_{\text{E}}$, let $\bar\gamma_{\text{R}}=\alpha\bar\gamma_{\text{D}}\rightarrow\infty$ ($\alpha>0$ is some constant), the SOP under Case \uppercase\expandafter{\romannumeral1} can be approximated as
\begin{align}
\mathcal{P}_1\approx{\mathcal{M}}_1{\bar\gamma_{\text{D}}^{-1}},\label{Theorem_Case1_2_Exp}
\end{align}
where ${\mathcal{M}}_1=\left(2^{\mathcal{R}}-1\right)\left(1+1/\alpha\right)+2^{\mathcal{R}}\bar\gamma_{\text{E}}>0$.
\end{theorem}
\vspace{-5pt}
\begin{IEEEproof}
When $\bar\gamma_{\text{E}}$ is fixed and $\bar\gamma_{\text{R}}=\alpha\bar\gamma_{\text{D}}\rightarrow\infty$, we have $\frac{2^{\mathcal{R}}-1}{\bar\gamma_{\text{R}}}\rightarrow0$, $\frac{2^{\mathcal{R}}-1}{\bar\gamma_{\text{D}}}\rightarrow0$, and $\frac{2^{\mathcal{R}}}{\bar\gamma_{\text{D}}}{\bar\gamma_{\text{E}}}\rightarrow0$. By substituting the Taylor expansions $\lim_{x\rightarrow0}{\rm{e}}^{-x}\approx1-x$ and $\lim_{x\rightarrow0}\frac{1}{1+x}=1-x$ into \eqref{Theorem_Case1_1_Exp}, we can get the high-SNR approximation in \eqref{Theorem_Case1_2_Exp} by neglecting the higher order terms.
\end{IEEEproof}
\vspace{-5pt}
\begin{remark}
The results in Theorem \ref{Theorem_Case1_2} suggest that $\mathcal{P}_1$ in the first scenario converges to $0$ in the high-SNR region and its rate of convergence (ROC) equals the rate of ${\mathcal{M}}{\bar\gamma_{\text{D}}^{-1}}$ converging to $0$. This fact means that the secrecy diversity order of $\mathcal{P}_1$ in the first scenario is given by $1$.
\end{remark}
\vspace{-5pt}
We then turn to the second scenario and conclude the following theorem.
\vspace{-5pt}
\begin{theorem}\label{Theorem_Case1_3}
Let $\bar\gamma_{\text{R}}=\alpha\bar\gamma_{\text{D}}=\beta\bar\gamma_{\text{E}}\rightarrow\infty$ ($\alpha>0$ and $\beta>0$ are some constants), the SOP under Case \uppercase\expandafter{\romannumeral1} can be approximated as
\begin{align}\label{Theorem_Case1_3_Exp}
\mathcal{P}_1\approx\mathcal{P}_1^{\lim}+{\hat{\mathcal{M}}_1}{\bar\gamma_{\text{D}}^{-1}},
\end{align}
where $\mathcal{P}_1^{\lim}=1-\frac{1}{1+\frac{\alpha}{\beta}{2^{\mathcal{R}}}}$ and ${\hat{\mathcal{M}}_1}=\frac{\left(2^{\mathcal{R}}-1\right)\left(1+1/\alpha\right)}{1+\frac{\alpha}{\beta}{2^{\mathcal{R}}}}$.
\end{theorem}
\vspace{-5pt}
\begin{IEEEproof}
Note that $\lim_{\bar\gamma_{\text{R}}\rightarrow\infty}{{\rm{e}}^{-\frac{2^{\mathcal{R}}-1}{\bar\gamma_{\text{R}}}}}=1$, $\lim_{\bar\gamma_{\text{D}}\rightarrow\infty}{{\rm{e}}^{-\frac{2^{\mathcal{R}}-1}{\bar\gamma_{\text{D}}}}}=1$, and $\frac{1}{1+\frac{2^{\mathcal{R}}}{\bar\gamma_{\text{D}}}{\bar\gamma_{\text{E}}}}=
\frac{1}{1+2^{\mathcal{R}}\frac{\alpha}{\beta}}$. It follows that $\lim_{\bar\gamma_{\text{R}}=\alpha\bar\gamma_{\text{D}}=\beta\bar\gamma_{\text{E}}\rightarrow\infty}{\mathcal{P}}_1=\mathcal{P}_1^{\lim}$. Besides, ${\mathcal{P}}_1-\mathcal{P}_1^{\lim}=\frac{1}{1+2^{\mathcal{R}}\frac{\alpha}{\beta}}\left(1-{\rm{e}}^{-\frac{2^{\mathcal{R}}-1}{\bar\gamma_{\text{R}}}-\frac{2^{\mathcal{R}}-1}{\bar\gamma_{\text{D}}}}\right)$. Using the Taylor expansion $\lim_{x\rightarrow0}{\rm{e}}^{-x}\approx1-x$, we can get the high-SNR approximation in \eqref{Theorem_Case1_3_Exp} after some basic manipulations.
\end{IEEEproof}
\vspace{-5pt}
\begin{remark}
The results in Theorem \ref{Theorem_Case1_3} suggest that $\mathcal{P}_1$ in the second scenario converges to $\mathcal{P}_1^{\lim}>0$ in the high-SNR region and its ROC equals the rate of ${\mathcal{M}_1}{\bar\gamma_{\text{D}}^{-1}}$ converging to $0$. This fact means that the secrecy diversity order of $\mathcal{P}_1$ in the second scenario is given by $1$.
\end{remark}
\vspace{-5pt}
\vspace{-5pt}
\begin{remark}
The SOP in the first scenario converges to zero in the high-SNR regime, whereas the SOP in the second scenario converges to a positive constant in the high-SNR regime.
\end{remark}
\vspace{-5pt}
\subsection{Case \uppercase\expandafter{\romannumeral2}}\label{sec4b}
Particularly, the following theorem gives the exact expression of the SOP.
\vspace{-5pt}
\begin{theorem}\label{Theorem_Case2_1}
The SOP under Case \uppercase\expandafter{\romannumeral2} is given as
\begin{align}
\mathcal{P}_2=1-{\rm{e}}^{-\frac{2^{\mathcal{R}}-1}{\bar\gamma_{\text{D}}}-\frac{2^{\mathcal{R}}-1}{\bar\gamma_{\text{R}}}}
\frac{1}{1+\frac{2^{\mathcal{R}}}{\bar\gamma_{\text{R}}}{\bar\gamma_{\text{E}}}}
\end{align}
\end{theorem}
\vspace{-5pt}
\begin{IEEEproof}
Similar to the proof of Theorem \ref{Theorem_Case1_1}.
\end{IEEEproof}
In the sequel, we perform high-SNR analyses to the SOP for gleaning more system insights.
\vspace{-5pt}
\begin{theorem}\label{Theorem_Case2_2}
For fixed $\bar\gamma_{\text{E}}$, let $\bar\gamma_{\text{R}}=\alpha\bar\gamma_{\text{D}}\rightarrow\infty$ ($\alpha>0$ is some constant), the SOP under Case \uppercase\expandafter{\romannumeral2} can be approximated as
\begin{align}
\mathcal{P}_2\approx{\mathcal{M}}_2{\bar\gamma_{\text{D}}^{-1}},\label{Theorem_Case2_2_Exp}
\end{align}
where ${\mathcal{M}}_2=\left(2^{\mathcal{R}}-1\right)\left(1+1/\alpha\right)+2^{\mathcal{R}}\bar\gamma_{\text{E}}/\alpha>0$.
\end{theorem}
\vspace{-5pt}
\begin{IEEEproof}
Similar to the proof of Theorem \ref{Theorem_Case1_2}.
\end{IEEEproof}
\vspace{-5pt}
\begin{theorem}\label{Theorem_Case2_3}
Let $\bar\gamma_{\text{R}}=\alpha\bar\gamma_{\text{D}}=\beta\bar\gamma_{\text{E}}\rightarrow\infty$ ($\alpha>0$ and $\beta>0$ are some constants), the SOP under Case \uppercase\expandafter{\romannumeral2} can be approximated as
\begin{equation}
\mathcal{P}_2\approx\mathcal{P}_2^{\lim}+{\hat{\mathcal{M}}_2}{\bar\gamma_{\text{D}}^{-1}},
\end{equation}
where $\mathcal{P}_2^{\lim}=1-\frac{1}{1+\frac{1}{\beta}{2^{\mathcal{R}}}}$ and $\hat{\mathcal{M}}_2=\frac{\left(2^{\mathcal{R}}-1\right)\left(1+1/\alpha\right)}{1+\frac{1}{\beta}{2^{\mathcal{R}}}}$.
\end{theorem}
\vspace{-5pt}
\begin{IEEEproof}
Similar to the proof of Theorem \ref{Theorem_Case1_3}.
\end{IEEEproof}
\vspace{-5pt}
\begin{remark}
The results in Theorem \ref{Theorem_Case2_2} and Theorem \ref{Theorem_Case2_3} suggest that the secrecy diversity orders of Case \uppercase\expandafter{\romannumeral2} in both the first and the second scenarios are given by 1.
\end{remark}
\vspace{-5pt}
\subsection{Case \uppercase\expandafter{\romannumeral3}}\label{sec4c}
Particularly, the following theorem gives the exact expression of the SOP.
\vspace{-5pt}
\begin{theorem}\label{Theorem_Case3_1}
The SOP under Case \uppercase\expandafter{\romannumeral3} is given as
\begin{align}
\mathcal{P}_3=1-{\rm{e}}^{-\frac{2^{\mathcal{R}}-1}{\bar\gamma_{\text{D}}}-\frac{2^{\mathcal{R}}-1}{\bar\gamma_{\text{R}}}}
\frac{1}{1+\frac{2^{\mathcal{R}}}{\bar\gamma_{\text{D}}}{\bar\gamma_{\text{E}}}}
\frac{1}{1+\frac{2^{\mathcal{R}}}{\bar\gamma_{\text{R}}}{\bar\gamma_{\text{E}}}}.
\end{align}
\end{theorem}
\vspace{-5pt}
\begin{IEEEproof}
Similar to the proof of Theorem \ref{Theorem_Case1_1}.
\end{IEEEproof}
Moreover, the high-SNR behaviours of $\mathcal{P}_3$ are described in the following theorems.
\vspace{-5pt}
\begin{theorem}\label{Theorem_Case3_2}
For fixed $\bar\gamma_{\text{E}}$, let $\bar\gamma_{\text{R}}=\alpha\bar\gamma_{\text{D}}\rightarrow\infty$ ($\alpha>0$ is some constant), the SOP under Case \uppercase\expandafter{\romannumeral3} can be approximated as
\begin{align}
\mathcal{P}_3\approx{\mathcal{M}}_3{\bar\gamma_{\text{D}}^{-1}}.\label{Theorem_Case3_2_Exp}
\end{align}
where ${\mathcal{M}}_3=\left(2^{\mathcal{R}}-1\right)\left(1+1/\alpha\right)+2^{\mathcal{R}}\bar\gamma_{\text{E}}\left(1+1/\alpha\right)>0$.
\end{theorem}
\vspace{-5pt}
\begin{IEEEproof}
Similar to the proof of Theorem \ref{Theorem_Case1_2}.
\end{IEEEproof}
\vspace{-5pt}
\begin{theorem}\label{Theorem_Case3_3}
Let $\bar\gamma_{\text{R}}=\alpha\bar\gamma_{\text{D}}=\beta\bar\gamma_{\text{E}}\rightarrow\infty$ ($\alpha>0$ and $\beta>0$ are some constants), the SOP under Case \uppercase\expandafter{\romannumeral3} can be approximated as
\begin{equation}
\mathcal{P}_3\approx\mathcal{P}_3^{\lim}+{\hat{\mathcal{M}}_3}{\bar\gamma_{\text{D}}^{-1}},
\end{equation}
where $\mathcal{P}_3^{\lim}=1-\frac{\beta}{\beta+{\alpha}{2^{\mathcal{R}}}}\frac{\beta}{\beta+{2^{\mathcal{R}}}}$ and ${\hat{\mathcal{M}}_3}=\frac{\left(2^{\mathcal{R}}-1\right)\left(1+1/\alpha\right)}{\left(1+\frac{\alpha}{\beta}{2^{\mathcal{R}}}\right)
\left(1+\frac{1}{\beta}{2^{\mathcal{R}}}\right)}$.
\end{theorem}
\vspace{-5pt}
\begin{IEEEproof}
Similar to the proof of Theorem \ref{Theorem_Case1_3}.
\end{IEEEproof}
\vspace{-5pt}
\begin{remark}
The results in Theorem \ref{Theorem_Case3_2} and Theorem \ref{Theorem_Case3_3} suggest that the secrecy diversity orders of Case \uppercase\expandafter{\romannumeral3} in both the first and the second scenarios are given by 1.
\end{remark}
\vspace{-5pt}
\subsection{Summary of the Results}\label{sec4d}
We now summarize all the analytical results in Table \ref{Table1} on the top of this page, where ``Limit'' and ``Order'' represent the limitation of the SOP in the high-SNR regime and the secrecy diversity order, respectively. As shown in this table, the secrecy diversity order of the considered secure RF system is given by $1$ for all the discussed cases and scenarios. Furthermore, taken Theorems \ref{Theorem_Case1_1}, \ref{Theorem_Case2_1}, and \ref{Theorem_Case3_1} together, we conclude the following results.
\vspace{-5pt}
\begin{remark}\label{Remark6}
When $\bar\gamma_{\text{D}}>\bar\gamma_{\text{R}}$, we have $\mathcal{P}_3>\mathcal{P}_1>\mathcal{P}_2$. When $\bar\gamma_{\text{D}}\leq\bar\gamma_{\text{R}}$, we have $\mathcal{P}_3>\mathcal{P}_2\geq\mathcal{P}_1$. These results indicate that the SOP in Case \uppercase\expandafter{\romannumeral3} is higher than that in either Case \uppercase\expandafter{\romannumeral1} or Case \uppercase\expandafter{\romannumeral2}. The reason lies in that the eavesdropper in Case \uppercase\expandafter{\romannumeral3} can overhear the confidential message from both S and R.
\end{remark}
\vspace{-5pt}
By Theorems \ref{Theorem_Case1_2}, \ref{Theorem_Case2_2}, and \ref{Theorem_Case3_2}, we get the result as follows.
\vspace{-5pt}
\begin{remark}\label{Remark7}
When $\alpha>1$, we have ${\mathcal{M}}_3>{\mathcal{M}}_1>{\mathcal{M}}_2$. When $\alpha\leq1$, we have ${\mathcal{M}}_3>{\mathcal{M}}_2>{\mathcal{M}}_1$. These results indicate that under the first scenario, the ROC of the SOP in Case \uppercase\expandafter{\romannumeral3} is slower than that in either Case \uppercase\expandafter{\romannumeral1} or Case \uppercase\expandafter{\romannumeral2}.
\end{remark}
\vspace{-5pt}
Additionally, taking Theorems \ref{Theorem_Case1_3}, \ref{Theorem_Case2_3}, and \ref{Theorem_Case3_3} together, we arrive at the result as follows.
\vspace{-5pt}
\begin{remark}\label{Remark8}
When $\alpha>1$, we have $\mathcal{P}_3^{\lim}>\mathcal{P}_1^{\lim}>\mathcal{P}_2^{\lim}$. When $\alpha\leq1$, we have $\mathcal{P}_3^{\lim}>\mathcal{P}_2^{\lim}\geq\mathcal{P}_1^{\lim}$.
\end{remark}
\vspace{-5pt}
\vspace{-5pt}
\begin{remark}\label{Remark9}
When $\alpha>1$, we have $\hat{\mathcal{M}}_2>\hat{\mathcal{M}}_1>\hat{\mathcal{M}}_3$. When $\alpha\leq1$, we have $\hat{\mathcal{M}}_1\geq\hat{\mathcal{M}}_2>\hat{\mathcal{M}}_3$. These results indicate that under the second scenario, the ROC of the SOP in Case \uppercase\expandafter{\romannumeral3} is faster than that in either Case \uppercase\expandafter{\romannumeral1} or Case \uppercase\expandafter{\romannumeral2}.
\end{remark}
\vspace{-5pt}

\begin{table}[!t]
\centering
\caption{Summary of the Results.}
\begin{tabular}{|c|cc|cc|}
\hline
Condition & \multicolumn{2}{c|}{$\bar\gamma_{\text{R}}=\alpha\bar\gamma_{\text{D}}\rightarrow\infty$}          & \multicolumn{2}{c|}{$\bar\gamma_{\text{R}}=\alpha\bar\gamma_{\text{D}}=\beta\bar\gamma_{\text{E}}\rightarrow\infty$}          \\ \hline
Metric    & \multicolumn{1}{c|}{Limit} & Order & \multicolumn{1}{c|}{Limit} & Order \\ \hline
Case \uppercase\expandafter{\romannumeral1}    & \multicolumn{1}{c|}{$0$}    & $1$     & \multicolumn{1}{c|}{$\mathcal{P}_1^{\lim}$}    & $1$     \\ \hline
Case \uppercase\expandafter{\romannumeral2}    & \multicolumn{1}{c|}{$0$}    & $1$     & \multicolumn{1}{c|}{$\mathcal{P}_2^{\lim}$}    & $1$     \\ \hline
Case \uppercase\expandafter{\romannumeral3}    & \multicolumn{1}{c|}{$0$}    & $1$     & \multicolumn{1}{c|}{$\mathcal{P}_3^{\lim}$}    & $1$     \\ \hline
\end{tabular}
\label{Table1}
\end{table}

\section{Numerical Results}\label{sec5}
In this section, numerical results will be used to demonstrate the secrecy performance of DF relay and also verify the accuracy of the developed analytical results.

\begin{figure}[!t]
\centering
\setlength{\abovecaptionskip}{0pt}
\includegraphics[width=0.4\textwidth]{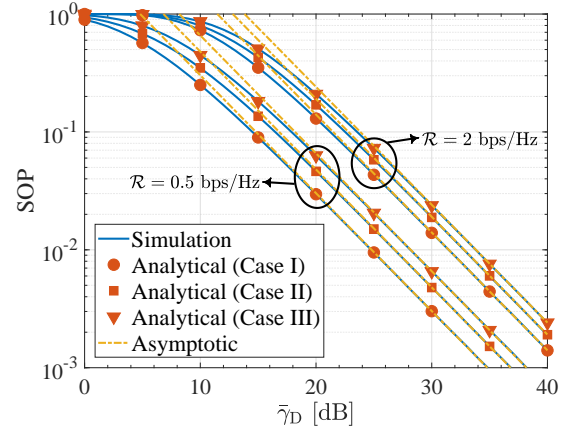}
\caption{SOP with ${\bar\gamma}_{\text{R}}=0.5{\bar\gamma}_{\text{D}}$ and ${\bar\gamma}_{\text{E}}=1$ dB.}
\label{Figure1}
\end{figure}

To confirm the accuracy of the analytical results developed in Theorems \ref{Theorem_Case1_1}, \ref{Theorem_Case2_1}, and \ref{Theorem_Case3_1}, we plot the SOPs of the three cases in terms of the SNR $\bar\gamma_{\text{D}}$ in {\figurename} {\ref{Figure1}} and compare the analytical results with the simulated results. As observed, the analytical results fit well with the simulations. Another observation is that the SOPs of the three cases satisfy $\mathcal{P}_3>\mathcal{P}_2>\mathcal{P}_1$, which is consistent with the conclusion drawn in Remark \ref{Remark6}. Furthermore, {\figurename} {\ref{Figure1}} also illustrates the high-SNR behaviour of the SOP under the first scenario, namely fixed $\bar\gamma_{\text{E}}$ and $\bar\gamma_{\text{R}}=\alpha\bar\gamma_{\text{D}}\rightarrow\infty$. Particularly, the asymptotic results are calculated by \eqref{Theorem_Case1_2_Exp}, \eqref{Theorem_Case2_2_Exp}, and \eqref{Theorem_Case3_2_Exp}. It can be seen from {\figurename} {\ref{Figure1}} that the derived asymptotic results track the numerical results accurately. Moreover, as can be seen from this graph, the SOP of Case \uppercase\expandafter{\romannumeral3} yields a slower ROC than the SOPs of the other two cases, which agrees with the conclusion in Remark \ref{Remark7}.

\begin{figure}[!t]
\centering
\setlength{\abovecaptionskip}{0pt}
\includegraphics[width=0.4\textwidth]{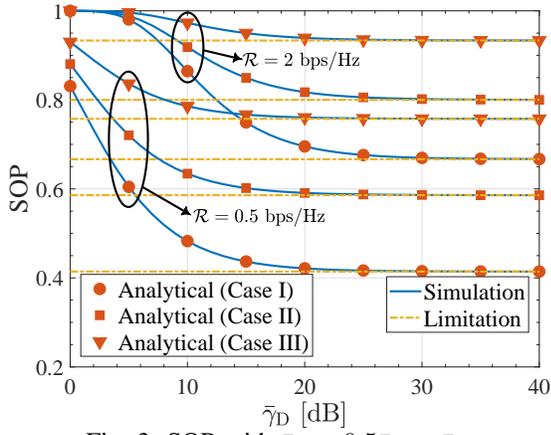}
\caption{SOP with ${\bar\gamma}_{\text{R}}=0.5{\bar\gamma}_{\text{D}}={\bar\gamma}_{\text{E}}$.}
\label{Figure2}
\end{figure}

\begin{figure}[!t]
\centering
\setlength{\abovecaptionskip}{0pt}
\includegraphics[width=0.4\textwidth]{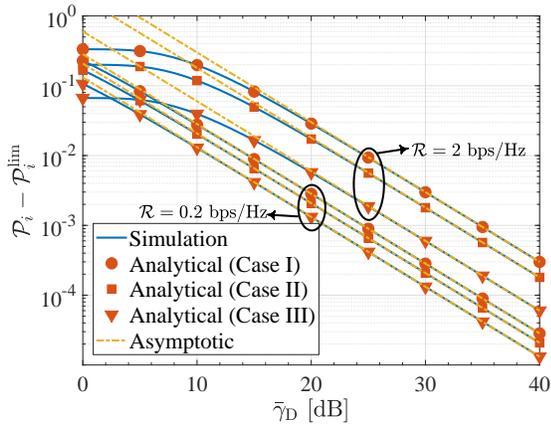}
\caption{ROC of the SOP with ${\bar\gamma}_{\text{R}}=0.5{\bar\gamma}_{\text{D}}={\bar\gamma}_{\text{E}}$.}
\label{Figure3}
\end{figure}

{\figurename} {\ref{Figure2}} plots the high-SNR SOPs versus the SNR $\bar\gamma_{\text{D}}$ for the considered three cases under the second scenario, namely $\bar\gamma_{\text{R}}=\alpha\bar\gamma_{\text{D}}=\beta\bar\gamma_{\text{E}}\rightarrow\infty$. As can be seen from this graph, with the increment of the SNR, the SOP converges to its limitation, i.e., $\lim_{\bar\gamma_{\text{D}}\rightarrow\infty}{\mathcal{P}_i}=\mathcal{P}_i^{\lim}$ ($i=1,2,3$), which supports the results in Theorems \ref{Theorem_Case1_3}, \ref{Theorem_Case2_3}, and \ref{Theorem_Case3_3}. Here, we set $\alpha=0.5<1$. Based on Remark \ref{Remark8}, we have $\mathcal{P}_3^{\lim}>\mathcal{P}_2^{\lim}\geq\mathcal{P}_1^{\lim}$ theoretically, which is consistent with the observation from {\figurename} {\ref{Figure2}}. To show the ROC of the SOP in the second scenario, we plot $\mathcal{P}_i-\mathcal{P}_i^{\lim}$ as a function of $\bar\gamma_{\text{D}}$ in {\figurename} {\ref{Figure3}}, where the analytical results are calculated by $\hat{\mathcal{M}}_i\bar\gamma_{\text{D}}^{-1}$. As shown, the derived asymptotic results match the simulated results perfectly in the high-SNR region. Besides, it can be seen from this graph that $\hat{\mathcal{M}}_3<\hat{\mathcal{M}}_2<\hat{\mathcal{M}}_1$, which agrees with the conclusion drawn in Remark \ref{Remark9}.

\section{Conclusion}\label{sec6}
This paper has studied the PHY security in DF relay systems. Rigorous derivations have been provided to derive new expressions of the secrecy capacity. On this condition, closed-form analytical results have been developed to characterize the explicit and asymptotic behaviours of the SOP. It was found that the SOP converges to some constant in the high-SNR regime and the its ROC is determined by the SDO.

\end{document}